\documentclass[pra,twocolumn,tightenlines,showpacs,nofootinbib,superscriptaddress]{revtex4-1}
\usepackage{bm,dcolumn,amsmath,graphicx,amsfonts,amssymb,longtable}

\usepackage{hyperref} 
\usepackage{xcolor}
\definecolor{blue}{rgb}{0.,0.,0.8}   
\hypersetup{ 
	colorlinks,
	linkcolor={blue},
	citecolor={blue},
	urlcolor={blue}
}

\newcommand{\bra}[1]{\ensuremath{\langle #1|}}	
\newcommand{\ket}[1]{\ensuremath{|#1\rangle}}	
\renewcommand{\v}[1]{\ensuremath{\boldsymbol{#1}}}		
\newcommand{\E}[1]{\ensuremath{\times10^{#1}}}	
\newcommand{\Exp}[1]{\ensuremath{e^{#1}}}	
\newcommand{\g}{\ensuremath{\gamma}} 
\newcommand{\el}[1]{\ensuremath{{}^{#1}}}

\begin{document} 
\title{Calculations of the atomic structure for the low-lying states of actinium}
\author{V. A. Dzuba}\email[]{v.dzuba@unsw.edu.au}
\author{V. V. Flambaum}\email[]{v.flambaum@unsw.edu.au}
\affiliation{School of Physics, University of New South Wales, Sydney 2052, Australia}
\author{B. M. Roberts}\email[]{b.roberts@uq.edu.au}
\altaffiliation[present address:~]{School of Mathematics and Physics, the University of Queensland, Brisbane, QLD 4072, Australia}
\affiliation{SYRTE, Observatoire de Paris, Universit\'e PSL, CNRS, Sorbonne Universit\'e, LNE, 61 avenue de l'Observatoire 75014 Paris, France}
\date{ \today }

\begin{abstract}
We employ a technique that combines the configuration interaction method with the singles-doubles coupled-cluster method to perform calculation of the 
energy levels, transition amplitudes, lifetimes, $g$-factors, and magnetic dipole and electric quadrupole hyperfine structure constants for 
many low-lying states of neutral actinium.
We find very good agreement with existing experimental energy levels and make accurate predictions for missing levels.
It has been noted that some of the levels have been previously misidentified; our analysis supports this claim.
If spectroscopy is performed with actinium-225, our calculations will lead to values for nuclear structure constants. 
The accuracy of this can be constrained by comparing with actinium-227.
\end{abstract}
\maketitle

\section{Introduction}

Actinium is a radioactive element with atomic number $Z=89$, and has three valence electrons above a radon-like core.
The most stable isotope, \el{227}Ac, has a half-life of 28 yr; the nearest even isotopes, \el{226,228}Ac, have lifetimes on the order of hours.
Its complicated electron structure and short half-life have made both theoretical and experimental investigations of its atomic properties quite difficult.
The data for the electron spectrum of Ac presented in the NIST database \cite{NIST} miss many levels and there are indications that some of the data are not accurate~\cite{Ferrer2017}, see also Refs.~\cite{Urer2012,Raeder2015,Granados2017}.

Several isotopes of actinium have extensive use in medical applications.
In particular, \el{225}Ac, an alpha-emitter with a 10 day half-life, is a very promising. The emitted radiation is sufficiently energetic to destroy cancer cells, but has a range short enough to be essentially contained and not damage nearby cells.
Work on producing and studying \el{225}Ac is ongong at the 
Los Alamos National Laboratory~\cite{Weidner2012a,Weidner2012}, 
Brookhaven National Laboratory~\cite{Fitzsimmons2018}, 
Leuven~\cite{Granados2017} and other centres.
The trapping and transport of these radioactive isotopes, however, is rather challenging.
A detailed knowledge of the energy levels and transition amplitudes is extremely important for the development and testing of trapping and cooling schemes.

We perform accurate calculations for neutral actinium using a technique based on the combination of the configuration interaction method with the linearized single-double--coupled-cluster method \cite{DzubaSDCI2014}.
We calculate the energy levels and $g$-factors for many of the low-lying states, including predictions for many previously unidentified levels, and electric dipole transition amplitudes between several of the lowest states.
Our calculations extend significantly beyond existing calculations (e.g.~Ref.~\cite{Urer2012,Eliav1998}), and are likely accurate to the few-percent level for the energy levels, and to around 10\% for most transition matrix elements.
Newer experimental work may indicate that some of the levels may have been previously misidentified (e.g.,~\cite{Ferrer2017,Raeder,Rossnagel2012,Raeder2013,Raeder2015}). 
Our calculations add weight to this claim, agreeing well with the newer measurements.
Actinium may also be of interest to studies of atomic parity and time-reversal violation \cite{RobertsActinides2013,*RobertsCL2014} (see also Ref.~\cite{RobertsReview2015}).
High-precision atomic structure calculations will be required for the interpretation of experimental results in this case.

\section{Atomic Structure Calculations}\label{sec:methods}

To perform the calculations, we use a method based on the combination of the configuration interaction (CI) method with the linearized single-double--coupled-cluster method (SD), as developed in Ref.~\cite{DzubaSDCI2014}. 
This method is similar to that developed in Ref.~\cite{Safronova2009a}, and good agreement between the two approaches has been demonstrated \cite{DzubaSDCI2014}.
This technique has proven to be effective and very accurate for few-valence-electron systems 
\cite{Safronova2009a,Safronova2013c,Safronova2013a,SafronovaTh2014,DzubaSDCI2014,Ginges2015}.

This method is similar to the combination of many-body perturbation theory (MBPT) with the CI method, as developed in Ref.~\cite{DzubaCIMBPT1996}.
Such a method has been used widely for a large number of applications and has been proven to be both efficient and accurate for few-valence atoms and ions.
The SD technique allows one to accurately take into account the core--valence and core-core electron correlations, while the CI method accounts for the valence--valence correlations.
The CI+SD method includes a more accurate treatment of the core--valence correlations, as well as the screening (by the core electrons) of the
Coulomb interaction between valence electrons compared to the CI+MBPT technique \cite{DzubaSDCI2014}.

We perform calculations in the $V^{N-M}$ approximation~\cite{DzubaVN-M2005}.
The calculations start with the Dirac-Fock procedure for the Ac~IV ion with all $M=3$ valence electrons removed. 
The single-electron basis states for valence electrons are calculated in the field of the Ac~IV ion. 
Removal of the valence electrons has very little effect on the core and thus on the CI Hamiltonian. 
On the other hand, the use of the $V^{N-M}$ approximation greately simlifies the calculation of the core-valence correlations
(for more details see Refs.~\cite{DzubaVN-M2005,Dzuba2005a,Dzuba2007}). 
The $V^{N-M}$ approximation is widely used in atomic calculations producing very accurate results (see, e.g.,~\cite{DzubaEDM2009,DzubaWeakRelations2011,KozlovTlEDM2012,Diaz2013,Imanbaeva2017}).
Use of the $V^{N-3}$ approximation has a clear and simple justification. Three valence electrons produce charge density which is mainly localised outside the electron core. As is well known from the electrodynamics, a homogeneously distributed charge on the surface of a sphere does not produce electric field inside; i.e., the potential inside is constant. Such a constant potential does not affect the wave functions of the core electrons or the contribution to the electrostatic potential produced by these core electrons. The core electron Hartree-Fock Green's function is also not affected. As a result, the perturbation theory approach for the core-valence interaction is sufficiently accurate and significantly simpler in the $V^{N-3}$ starting approximation

In the SD method, the many-body wave function is expressed as an expansion that contains all single and double excitations from the Hartree-Fock reference wave function, see, e.g., Ref.~\cite{Blundell1989}.
The coefficients of the expansion are found by solving the set of SD equations.
First, the SD equations are solved self consistently for the core electrons to determine the core excitation coefficients. 
Then, the SD equations are solved for valence states. This a complete procedure for the case of a single-valence electron.
For systems with more than one valence electron, however, the interactions between the valence electrons must also be taken into account. This is done with the use of the CI technique (see Ref.~\cite{DzubaSDCI2014} for details).

In the CI method, the many-electron wave function for a valence state is expressed as a linear combination of Slater determinants,
constructed from single-electron basis states 
$\Psi_j = \sum_i c_i \psi_{ji}$.
The Slater determinants are constructed from configurations, which in turn are obtained by all single and double, and selected triple and quadrupole excitations from the reference configuration.
We use the B-spline technique~\cite{JohnsonBspl1986} to make a single electron basis for Slater determinants $\psi_{jl}$.

The effective CI Hamiltonian in the CI equation $\hat H^{\rm CI} \Psi_j = E_j \Psi_j$ 
can be expressed as the sum of single-particle and two-particle operators,
\begin{equation}
\hat H^{\rm CI} = \sum_{i=1}^3 \hat h_1(\v{r}_i) + \sum_{i<j}\hat h_2(\v{r}_i,\v{r}_j),
\label{eq:HCI}
\end{equation}
where the summation runs over the valence electrons (Ac has three valence electrons).
The one- and two-electron parts of the Hamiltonian are given by
\begin{equation}
 \hat{h}_1(\v{r}) = c\v{\alpha}\cdot\v{p}+ (\g^0-1)c^2+\hat{V}^{\rm nuc}+\hat V^{N-3} + \hat \Sigma_1,
\label{eq:HCI-1}
\end{equation}
and 
\begin{equation}
\hat h_2(\v{r}_1,\v{r}_2) = r_{12}^{-1} + \hat\Sigma_2(\v{r}_1,\v{r}_2),
\label{eq:HCI-2}
\end{equation}
respectively,
where $\g^0$ and $\v{\alpha}=\g^0\v{\g}$ are Dirac matrices, \v{p} is the relativistic (three-)momentum of the electron, $\hat{V}^{\rm nuc}$  is the nuclear potential (for which we use a Fermi distribution of nuclear charge),  $\hat V^{N-3}$ is the frozen Hartree-Fock potential created by the $N-3$ core electrons, and $r_{12}=\left|{\v{r}_1-\v{r}_2}\right|$.
The other operators, $\hat \Sigma_1$ and $\hat \Sigma_{2}$, are the correlation operators, which are determined from the SD equations. 
Without these, the above equations would correspond to the conventional CI method.
The single-particle operator $\hat \Sigma_1$ corresponds to the interaction of a valence electron with the atomic core. 
The two-particle operator, $\hat \Sigma_2$, represents the screening of the valence--valence Coulomb interaction by the core electrons.

\subsection{Breit and QED effects}

We also take into account two non-correlation corrections; those due to the Breit and radiative quantum electrodynamics (QED) effects.
The effective Breit Hamiltonian, which includes magnetic and retardation corrections (in the zero energy transfer limit) can be expressed
\begin{equation}
\hat h^{\rm B}(\v{r}_1,\v{r}_2)=\frac{-1}{2r_{12}}\left(\v{\alpha}_1\cdot\v{\alpha}_2+\frac{(\v{\alpha}_1\cdot\v{r}_{12})(\v{\alpha}_2\cdot\v{r}_{12})}{r_{12}^2}\right),
\end{equation}
where $\v{r}_{12}=\v{r}_1-\v{r}_2$.

We include radiative QED effects via the use of the so-called radiative potential method, developed in Ref.~\cite{FlambaumQED2005}.
The radiative potential has the form
\begin{equation}
V^{\rm rad}(r)= V_{\rm U}(r) + V_{\rm g}(r) + V_{\rm e}(r),
\end{equation}
where $V_{\rm U}$ is the Uehling potential (lowest order vacuum polarization correction),
and the self-energy corrections $V_{\rm g}$ and $V_{\rm e}$ are the potentials arising from the magnetic and electric electron form-factors, respectively.
The $V_{\rm U}$ and $V_{\rm e}$ terms can be added directly to the Hartree-Fock potential, while $V_{\rm g}$ leads to corrections to the Dirac equation (see Ref.~\cite{FlambaumQED2005}).
Relaxation is included for both the Breit and QED potentials by including them in the iterations of the Hartree-Fock equations.

\subsection{Matrix elements and core polarisation}

To calculate the matrix elements for the transition amplitudes we use the time-dependent Hartree-Fock (TDHF) method, as in Refs.~\cite{DzubaHFS1984,Dzuba1987jpb,Dzuba1987jpbRPA}.
In the presence of external fields, the Hartree-Fock core is polarized and its potential should be modified:
\begin{equation}\hat V^{N-3}\to\hat V^{N-3}+\delta V.\end{equation}
To calculate $\delta V$, the wave function is written 
as
\begin{equation}
\psi=\psi_0 + \eta\Exp{-i\omega t}+ \zeta\Exp{i\omega t},
\end{equation}
where $\psi_0$ is the unperturbed wave function, $\eta$ and $\zeta$ are corrections due to the external field, and $\omega$ is the frequency of the external field (frequency of the transition).


\begin{longtable*}{D{.}{2}{2}ll D{,}{}{9.6}rD{,}{}{7.4} D{.}{.}{1.4}}
\caption{
Calculated excitation energies and $g$-factors for the lowest states of Ac,
and comparison with experiment ($\Delta = E_{\rm exp} - E_{\rm calc}$).
Where the term designations given in the NIST database \cite{NIST} differ from those determined in this work, the term symbol from Ref.~\cite{NIST} is given inside brackets following the experimental energy.} \label{tab:Ac-energy}\\
\hline\hline
\multicolumn{3}{c}{Level}&
\multicolumn{3}{c}{Energy (cm$^{-1}$)}&
\multicolumn{1}{c}{ }\\
\cline{1-3}\cline{4-6}
\multicolumn{1}{l}{$\ N$}& \multicolumn{1}{c}{Conf.~~~~~~~~~~}&
\multicolumn{1}{c}{Term}&
\multicolumn{1}{c}{$E_{\rm exp}$ \cite{NIST}}&
\multicolumn{1}{c}{$E_{\rm calc}$}&
\multicolumn{1}{c}{$\Delta$}&
\multicolumn{1}{c}{$g$}\\
\hline
\endfirsthead
\multicolumn{7}{l}%
{{\tablename\ \thetable{}}: continued from previous page} \\
\hline\hline
\multicolumn{1}{l}{$N$}& \multicolumn{1}{c}{Conf.~~~~~~~~~~}&
\multicolumn{1}{c}{Term}&
\multicolumn{1}{c}{$E_{\rm exp}$}&
\multicolumn{1}{c}{$E_{\rm calc}$}&
\multicolumn{1}{c}{$\Delta$}&
\multicolumn{1}{c}{$g$}\\
\hline
\endhead
\hline \multicolumn{7}{l}{{Continued on next page}} \\ \hline
\endfoot
\endlastfoot
1 & $7s^{2}6d$ & $^2$D$_{3/2}$ & 0 & 0 & 0 & 0.8001\\
2 & $7s^{2}6d$ & $^2$D$_{5/2}$ & 2231 & 2339 & -108 & 1.2002\\
3 & $7s^{2}7p$ & $^2$P$^o_{1/2}$ &  & 7565 &  & 0.6626\\
4 & $7s6d^{2}$ & $^4$F$_{3/2}$ & 9217 & 8989 & 228 & 0.4088\\
5 & $7s6d^{2}$ & $^4$F$_{5/2}$ & 9864 & 9288 & 576 & 1.0298\\
6 & $7s6d^{2}$ & $^4$F$_{7/2}$ & 10906 & 9974 & 932 & 1.2333\\
7 & $7s6d^{2}$ & $^4$F$_{9/2}$ & 12078 & 11726 & 352 & 1.3143\\
8 & $7s^{2}7p$ & $^2$P$^o_{3/2}$ &  & 12345 &  & 1.3332\\  
9 & $7s6d^{2}$ & $^4$P$_{1/2}$ &  & 12583 &  & 2.6295\\
10 & $7s6d^{2}$ & $^4$P$_{3/2}$ &  & 12847 &  & 1.6841\\
11 & $7s6d^{2}$ & $^4$D$_{5/2}$ &  & 13301 &  & 1.4320\\
12 & $7s7p6d$ & $^4$F$^o_{3/2}$ & 13713 & 13958 & -245 & 0.4778\\
13 & $7s6d^{2}$ & $^2$F$_{5/2}$ &  & 14810 &  & 0.9302\\
14 & $7s6d^{2}$ & $^2$D$_{3/2}$ &  & 14985 &  & 0.9489\\
15 & $7s7p6d$ & $^4$F$^o_{5/2}$ & 14941 & 15141 & -200 & 1.0722\\
16 & $7s6d^{2}$ & $^4$G$_{7/2}$ &  & 15746 &  & 0.9832\\
17 & $7s6d^{2}$ & $^2$P$_{1/2}$ &  & 16345 &  & 0.7371\\
18 & $7s7p6d$ & $^4$D$^o_{1/2}$ & 17200 & 17049 & 151 & 0.0698\\
19 & $7s6d^{2}$ & $^2$G$_{9/2}$ &  & 17198 &  & 1.1305\\
20 & $7s6d^{2}$ & $^4$D$_{5/2}$ &  & 17329 &  & 1.2931\\
21 & $7s7p6d$ & $^4$D$^o_{3/2}$ & 17736,~[^2{\rm D}] & 17612 & 124 & 1.1984\\
22 & $7s7p6d$ & $^4$F$^o_{7/2}$ & 17684 & 17715 & -31 & 1.2502\\
23 & $7s7p6d$ & $^2$D$^o_{5/2}$ & 17951 & 18108 & -157 & 1.2684\\
24 & $7s6d^{2}$ & $^4$G$_{7/2}$ &  & 18606 &  & 1.0530\\
25 & $7s7p6d$ & $^4$D$^o_{5/2}$ &  & 18747 &  & 1.3137\\
26 & $7s7p6d$ & $^2$D$^o_{3/2}$ & 19012,~[^4{\rm D}] & 18961 & 51 & 0.8590\\
27 & $7s6d^{2}$ & $^4$D$_{3/2}$ &  & 18980 &  & 1.2225\\
28 & $7s7p6d$ & $^4$D$^o_{7/2}$ &  & 20288 &  & 1.3858\\
29 & $7s7p6d$ & $^4$F$^o_{9/2}$ &  & 20640 &  & 1.3333\\
30 & $7s7p6d$ & $^4$P$^o_{1/2}$ & 22402,\tablenotemark[1] & 20669 & 1733 & 2.5826\\
31 & $7s7p6d$ & $^4$P$^o_{3/2}$ & 22801,\tablenotemark[2] & 20877 & 1924 & 1.5966\\
32 & $7s7p6d$ & $^4$F$^o_{5/2}$ & 21196,~[^4{\rm D}] & 21170 & 26 & 1.0394\\
33 & $7s6d^{2}$ & $^2$S$_{1/2}$ &  & 21918 &  & 1.9806\\
34 & $7s7p6d$ & $^2$D$^o_{5/2}$ & 23899,~[^4{\rm P}] & 22281 & 1618 & 1.2640\\%
35 & $7s7p6d$ & $^4$F$^o_{5/2}$ & 23917,~[^2{\rm F}] & 23379 & 538 & 0.9769\\
36 & $6d^{3}$ & $^4$F$_{3/2}$ &  & 23485 &  & 0.4253\\
37 & $6d^{3}$ & $^4$F$_{5/2}$ &  & 23657 &  & 1.0359\\
38 & $7s7p6d$ & $^2$D$^o_{3/2}$ &  & 23828 &  & 0.8534\\
39 & $7s7p6d$ & $^2$F$^o_{7/2}$ & 23476,~[^4{\rm D}] & 24015 & -539 & 1.1600\\
40 & $7s7p6d$ & $^2$F$^o_{7/2}$ & 24969 & 24532 & 437 & 1.1513\\
41 & $6d^{3}$ & $^4$F$_{7/2}$ &  & 24692 &  & 1.2114\\
42 & $7s7p6d$ & $^2$P$^o_{1/2}$ & 25729 & 25421 & 308 & 0.6628\\
43 & $6d^{3}$ & $^4$G$_{9/2}$ &  & 25738 &  & 1.2375\\
44 & $7s7p6d$ & ${}^2$P${}^o_{3/2}$ & 26066,~[^2{\rm D}] & 26235 & -168 & 1.2932\\
45 & $7s7p6d$ & $^4$F$_{5/2}^o$ & 26533,~[^2{\rm D}] & 26595 & -62 & 1.0389\\
46 & $7p6d^2$ & ${}^2$F${}^o_{5/2}$ & 26836 & 27029 & -192 & 0.7184\\
47 & $7s^{2}8s$ & $^2$S$_{1/2}$ &  & 27097 &  & 2.0134\\
48 & $6d^{3}$ & $^4$P$_{1/2}$ &  & 27353 &  & 2.4696\\
49 & $6d^3$ & $^2$P$_{3/2}$ &  & 27634 &  & 1.5303\\
50 & $7p6d^{2}$ & $^4$G$^o_{7/2}$ & 28568,~[^2{\rm F}] & 27966 & 602 & 0.9936\\
51 & $6d^{3}$ & $^2$G$_{7/2}$ &  & 28000 &  & 0.9210\\
52 & $7s7p6d$ & $^2$P$^o_{1/2}$ &  & 28156 &  & 0.6878\\
53 & $7s7p6d$ & ${}^2$P${}^o_{3/2}$ & 27010 & 28169 & -1159 & 1.3212\\
54 & $6d^3$ & $^4$P$_{5/2}$ &  & 28201 &  & 1.5024\\
55 & $6d^{3}$ & $^4$H$_{9/2}$ &  & 28402 &  & 1.0156\\
56 & $6d^3$ & $^4$D$_{3/2}$ &  & 28793 &  & 1.0036\\
57 & $7s^25f$ & ${}^2$F${}^o_{5/2}$ &  & 29063 &  & 0.8621\\
58 & $7p6d^{2}$ & $^4$G$^o_{9/2}$ &  & 29651 &  & 1.1646\\
59 & $7s7p^{2}$ & $^4$P$_{1/2}$ &  & 30727 &  & 2.5566\\
60 & $6d^3$ & $^2$D$_{5/2}$ &  & 30787 &  & 1.2283\\
61 & $7p6d^2$ & ${}^4$F${}^o_{3/2}$ & 30397,~[^2{\rm P}] & 30803 & -407 & 0.4273\\
62 & $7p6d^2$ & ${}^4$F${}^o_{5/2}$ & 31495,~[^4{\rm G}] & 31333 & 161 & 1.0242\\
63 & $7s^{2}5f$ & $^2$F$^o_{7/2}$ &  & 31366 &  & 1.1238\\
64 & $7s8s6d$ & $^2$D$_{3/2}$ &  & 31558 &  & 0.9099\\
65 & $7s^27d$ & $^2$D$_{5/2}$ &  & 31874 &  & 1.1915\\
66 & $7s8s6d$ & $^4$D$_{3/2}$ &  & 31937 &  & 1.1587\\
67 & $7p6d^{2}$ & $^2$G$^o_{9/2}$ &  & 32166 &  & 1.1172\\
68 & $7s7p^2$ & $^4$P$_{3/2}$ &  & 32495 &  & 1.6364\\
69 & $7s8s6d$ & $^4$D$_{5/2}$ &  & 32611 &  & 1.3124\\
70 & $7p6d^2$ & ${}^2$D${}^o_{3/2}$ &  & 32697 &  & 0.9900\\
71 & $7p6d^{2}$ & $^2$S$^o_{1/2}$ &  & 32747 &  & 1.5230\\ 
72 & $6d^3$ & $^4$F$_{5/2}$ &  & 32902 &  & 1.0395\\
73 & $7p6d^2$ & ${}^4$F${}^o_{5/2}$ &  & 32934 &  & 1.0626\\
74 & $6d^3$ & $^2$D$_{3/2}$ &  & 33365 &  & 0.9511\\
75 & $7p6d^2$ & ${}^4$D${}^o_{3/2}$ &  & 33506 &  & 1.1784\\
76 & $7s^29p$ & ${}^2$P${}^o_{3/2}$ &  & 33551 &  & 1.3310\\
77 & $7p6d^2$ & ${}^4$F${}^o_{5/2}$ &  & 33635 &  & 1.0619\\
78 & $7p6d^2$ & ${}^4$D${}^o_{3/2}$ &  & 34208 &  & 1.0352\\
79 & $7p6d^2$ & ${}^4$D${}^o_{5/2}$ &  & 34290 &  & 1.3224\\
80 & $6d^3$ & $^4$D$_{3/2}$ &  & 34409 &  & 1.0191\\
81 & $7s7p^2$ & $^4$D$_{5/2}$ &  & 34514 &  & 1.4652\\
82 & $7p6d^2$ & ${}^4$D${}^o_{5/2}$ &  & 35115 &  & 1.3015\\
83 & $7s8s6d$ & $^2$D$_{5/2}$ &  & 35290 &  & 1.2158\\
84 & $7p6d^2$ & ${}^4$D${}^o_{3/2}$ &  & 35461 &  & 1.1724\\
85 & $7s6d7d$ & $^4$G$_{5/2}$ &  & 36150 &  & 0.6868\\
86 & $7s6d7d$ & $^2$D$_{3/2}$ &  & 36218 &  & 0.8778\\
\hline\hline\\[-0.325cm]
\multicolumn{7}{l}{\tablenotemark[1]It is likely that this level does not originate from a transition to the ground  
}\\\multicolumn{7}{l}{
~ state but from the $^2$D$_{5/2}$ state at 2231\,cm$^{-1}$; the correct energy would then  
}\\\multicolumn{7}{l}{
~ be 24631\,cm$^{-1}$, with $J=3/2,\ 5/2,\ {\rm or }\ 7/2$~ \cite{Raeder}. Most likely, this    %
}\\\multicolumn{7}{l}{
~ corresponds to one of the levels denoted here as $N=38$, 39, or 40.}\\
\multicolumn{7}{l}{\tablenotemark[2]This level actually has $J=5/2$ \cite{Ferrer2017}, and most likely corresponds to the level
}\\\multicolumn{7}{l}{
~ denoted here as $N=34$.}
\end{longtable*}


Then, the set of TDHF equations
\begin{align}
(\hat h^{\rm HF} - \varepsilon_c - \omega)\eta_c &=-(\hat h_{\rm ext} + \delta V-\delta\varepsilon_c)\psi_c\\
(\hat h^{\rm HF} - \varepsilon_c + \omega)\zeta_c &=-(\hat h_{\rm ext}^\dagger + \delta V^\dagger-\delta\varepsilon_c)\psi_c,
\label{eq:tdhf}
\end{align}
are solved self-consistently for the core orbitals.
Here, the index $c$ denotes a state in the core, $\hat h_{\rm ext}$ is the operator of the external field interaction, and $\delta\varepsilon=\bra{\psi_0}\delta V\ket{\psi_0}$ is the correction to the energy due to the external field.

Core polarization is included into the calculation of the the matrix elements via a redefinition of the external field operators, e.g.,
\[\v{d}_{E1}\to\widetilde{\v{d}}_{E1} =\v{d}_{E1}+\delta V_{E1}, \]
where $\v{d}_{E1}=-e\v{r}$ is the operator of the $E1$ interaction, and $\delta V_{E1}$ is the correction to the Hartree-Fock core potential due to the action of the external $E1$ photon field.
This method is equivalent to the random phase approximation (RPA) method \cite{Johnson1989}, and includes core-polarization effects to all-orders.

\begin{table*}
\caption{\label{tab:hfs}
Magnetic dipole ($A$) and electric quadrupole ($B$) hyperfine structure constants for several states of Ac.
The values for $A$ are given in units of ${\frac{\mu}{I}{\rm MHz}}$, where $\mu$ is the nuclear magnetic dipole moment and $I$ is the nuclear spin, and $B$ are given in units of $\frac{Q}{\rm b}\,{\rm MHz}$, where $Q$ is the nuclear electric quadrupole moment and $b=1\E{-28}\,{\rm m}^2$.
For \el{227}Ac, $\mu=1.1$, $Q=1.7\,{\rm b}$, and $I=3/2$. 
} 
\begin{ruledtabular}
\begin{tabular}{llllD{,}{}{5.2}D{,}{}{5.2}D{.}{.}{3.2}D{.}{.}{3.2}}
\multicolumn{4}{c}{State}&
\multicolumn{2}{c}{${A}/{\frac{\mu}{I}{\rm MHz}}$}&
\multicolumn{2}{c}{$B/\frac{Q}{\rm b}\,{\rm MHz}$}\\
\cline{1-4}\cline{5-6}\cline{7-8}
\multicolumn{1}{c}{$N$}&
\multicolumn{2}{c}{Configuration}&
\multicolumn{1}{c}{Energy (cm$^{-1}$)}&
\multicolumn{1}{c}{Calc.}&
\multicolumn{1}{c}{Exp.}&
\multicolumn{1}{c}{Calc.}&
\multicolumn{1}{c}{Exp.}\\
\hline
1&$7s^2 6d $ &$^2$D$_{3/2} $&        0     	 	&-51,.71 	&70, 		& 372		&346	\\
2&$7s^2 6d $ &$^2$D$_{5/2} $&        2339     	& 211,.2 	&348,		& 464		&429	\\
34&$7s7p6d $& $^2$D$^o_{5/2}$&  22281 	& 2161,  	&			& 128 		&  	\\
42&$7s7p6d $ &$^2$P$^o_{1/2}$&  25421 	& 2176,    	&2692,		& 0			&0	\\  
44&$7s7p6d$  &$^2$P$^o_{3/2}$&  26235 	&-1391,     	&-1351,	& 54.2		&-15.6	\\
\end{tabular}
\end{ruledtabular}
\end{table*}

\section{Results and Discussion}\label{sec:results}

In Table~\ref{tab:Ac-energy}, we present our calculations of the energy levels and $g$-factors for several of the low-lying states for Ac.
Also presented are the accepted experimental values (taken from the NIST database \cite{NIST}), and the difference between these and our calculations (in the column $\Delta$).
For most levels, the agreement is better than a few percent.
In Ref.~\cite{Eliav1998}, calculations of excitation energies for three of the lowest states of Ac were performed using a coupled-cluster approach, including non-linear terms, and we find good agreement with these calculations.

As noted, it is possible that a number of these levels have been previously misidentified in the literature. 
The three largest deviations from experiment are likely explained by misidentification of the levels \cite{Raeder,Urer2012}, see footnotes of Table~\ref{tab:Ac-energy}. 
In Ref.~\cite{Urer2012}, the authors employed a multi-configuration Dirac-Fock calculation (including Breit and QED effects) to calculate the spectrum of the low-lying states of neutral actinium.
Our calculations agree reasonably with those in Ref.~\cite{Urer2012}, however, our work represents a significant improvement in accuracy.
This becomes more noticeable at higher energies.

Table \ref{tab:hfs} presents our calculated hyperfine structure constants, and comparison with experimental values.
The magnetic dipole constant $A$ is relatively unstable in the calculations for the ground state, while the electric quadrupole constant $B$ is comparatively stable. 
This is due primarily to the role of $d$-states. 
The direct contribution is small, because matrix elements for magnetic hyperfine structure are small for $d$-states; the value of $A$ comes from many-body corrections where $s$ and $p$ states play role. 
In contrast, $s$ and $p_{1/2}$ states do not contribute to $B$ (due to their too small total angular momentum $j$: $\bra{1/2}\hat Q\ket{1/2}=0$), but $d$-states contribute significantly.

Actually presented in the tables are $A/(\mu/I)$, and $B/Q$, where $\mu$ is the nuclear magnetic moment (in nuclear magnetons), $I$ is the nuclear spin, and $Q$ is the magnetic quadrupole moment (in barns). 
These numbers are presented since they are independent of the nuclear parameters (besides the effects of finite nuclear size, which are well below the assumed accuracy).
The corresponding experimental values are found using the known parameters for \el{227}Ac, $\mu=1.1$, $Q=1.7$, and $I=3/2$.
If measurements of the hyperfine structure are performed for \el{225}Ac~\cite{Raeder2015}, these calculations can be used to extract the nuclear parameters $\mu$ and $Q$. The accuracy of these predictions can be gauged from the comparison with \el{227}Ac.

In Table~\ref{tab:Ac-MEs}, we present calculations of the reduced matrix elements for the electric dipole ($E1$) amplitudes between several of the lowest states of Ac.
In order to control the accuracy of the calculations, we also performed all calculations using the CI+MBPT method, as developed in Ref.~\cite{DzubaCIMBPT1996}, which includes a set of dominating correlation diagrams to the second-order in perturbation theory.
By comparing the results of the CI+MBPT calculations with the all-order CI+SD calculations, a reasonable estimate of the uncertainty due to missed higher-order correlations can be formed.
For nearly all transitions, the difference between these calculations is less than 10\%.

The total decay probability for state $i$ is given {${\Gamma_i=\sum_j\gamma_{ij}}$}, where the summation is over all lower state $j$, and (considering only $E1$ transitions) the partial transition probability is (in atomic units)
\begin{equation}\label{eq:partial-rate}
\gamma_{ij} = \frac{4}{3}(\alpha\omega_{ij})^3\frac{|\bra{j}|\hat d_{E1}|\ket{i}|^2}{2J_i+1}.
\end{equation}
Here, $\bra{j}|\hat d_{E1}|\ket{i}$ is the reduced matrix element for the $E1$ amplitude,  $\omega_{ij}=E_i-E_j$, and $\alpha\approx1/137$ is the fine structure constant.
The lifetime for the state $i$ is then $1/\Gamma_i$.
To convert the rate from atomic units to ordinary units we should multiply the result in Eq.~(\ref{eq:partial-rate}) by  ($2 Ry/\hbar$). 
Correspondingly, to find the lifetime we should multiply $1/\Gamma$ in atomic units by $(\hbar/2 Ry) \approx 2.4189\E{-17}\,{\rm s}$.

In Table~\ref{tab:lifetimes}, we present our calculations of the lifetimes for several of the lowest-lying odd states.
The presented errors take into account the uncertainties in the calculated frequencies (where experimental values were not available), and the uncertainties in the $E1$ matrix elements.
For the higher states, the uncertainties also reflect the fact that only the dominating transitions were included.
To determine the uncertainties, we conservatively take the uncertainties in the frequencies and $E1$ amplitudes to be 10\% and 20\%, respectively.
We calculate the lifetime of the state
$7s^{2}7p$~$^2$P$^o_{ 1/2}$
(denoted as state 3 in Table~\ref{tab:Ac-energy}) to be
$4\E{-7}\,{\rm s}$,
with an uncertainty of about 50\% (which comes mostly from the calculation of the frequency).

\begin{table}
\footnotesize
  \caption{Electric dipole transition amplitudes (reduced matrix elements, in atomic units) between nine lowest odd states and all lower even states.\label{tab:Ac-MEs}
} 
\begin{ruledtabular}
  \begin{tabular}{llr c llld}
    \multicolumn{2}{c}{Odd States} &
    \multicolumn{3}{c}{$N_o - N_e$} &
    \multicolumn{2}{c}{Even States} &
    \multicolumn{1}{c}{Amplitude} \\
    \hline
    $7s^27p$ & $^2$P$^{\rm o}_{1/2}$ & 3 &-& 1 & $7s^26d$ & $^2$D$_{3/2}$ & 1.7639 \\
    $7s^27p$ & $^2$P$^{\rm o}_{3/2}$ & 8 &-& 1 & $7s^26d$ & $^2$D$_{3/2}$ & 0.5564 \\
                     &           & 8 &-& 2 & $7s^26d$ & $^2$D$_{5/2}$ & 2.7251 \\
                     &           & 8 &-& 4 & $7s6d^2$ & $^4$F$_{3/2}$ & 0.5812 \\
                     &           & 8 &-& 5 & $7s6d^2$ & $^4$F$_{5/2}$ & 0.7357 \\
    $7s7p6d$ & $^4$F$^{\rm o}_{3/2}$ &12 &-& 1 & $7s^26d$ & $^2$D$_{3/2}$ & 1.4222 \\
                     &           & 12 &-& 2 & $7s^26d$ & $^2$D$_{5/2}$ & 0.5030 \\
                     &           & 12 &-& 4 & $7s6d^2$ & $^4$F$_{3/2}$ & 2.5923 \\
                     &           & 12 &-& 5 & $7s6d^2$ & $^4$F$_{5/2}$ & 0.4852 \\
                     &           & 12 &-& 10 & $7s6d^2$ & $^4$P$_{3/2}$ & 0.1032 \\
    $7s7p6d$ & $^4$F$^{\rm o}_{5/2}$ &15 &-& 1 & $7s^26d$ & $^2$D$_{3/2}$ & 1.6269 \\
                     &           & 15 &-& 2 & $7s^26d$ & $^2$D$_{5/2}$ & 1.4982 \\
                     &           & 15 &-& 4 & $7s6d^2$ & $^4$F$_{3/2}$ & 1.2340 \\
                     &           & 15 &-& 5 & $7s6d^2$ & $^4$F$_{5/2}$ & 2.9311 \\
                     &           & 15 &-& 6 & $7s6d^2$ & $^4$F$_{7/2}$ & 0.0685 \\
                     &           & 15 &-& 10 & $7s6d^2$ & $^4$P$_{3/2}$ & 0.0316 \\
                     &           & 15 &-& 11 & $7s6d^2$ & $^4$D$_{5/2}$ & 0.3969 \\
                     &           & 15 &-& 13 & $7s6d^2$ & $^2$F$_{5/2}$ & 0.2210 \\
                     &           & 15 &-& 14 & $7s6d^2$ & $^2$D$_{3/2}$ & 0.4773 \\
    $7s7p6d$ & $^4$D$^{\rm o}_{1/2}$ &18 &-& 1 & $7s^26d$ & $^2$D$_{3/2}$ & 0.6240 \\
                     &           & 18 &-& 4 & $7s6d^2$ & $^4$F$_{3/2}$ & 3.1520 \\
                     &           & 18 &-& 9 & $7s6d^2$ & $^4$P$_{1/2}$ & 0.7208 \\
                     &           & 18 &-& 10 & $7s6d^2$ & $^4$P$_{3/2}$ & 0.1489 \\
                     &           & 18 &-& 14 & $7s6d^2$ & $^2$D$_{3/2}$ & 0.3500 \\
                     &           & 18 &-& 17 & $7s6d^2$ & $^2$P$_{1/2}$ & 0.2235 \\
    $7s7p6d$ & $^4$D$^{\rm o}_{3/2}$ &21 &-& 1 & $7s^26d$ & $^2$D$_{3/2}$ & 0.7445 \\
                     &           & 21 &-& 2 & $7s^26d$ & $^2$D$_{5/2}$ & 0.6392 \\
                     &           & 21 &-& 4 & $7s6d^2$ & $^4$F$_{3/2}$ & 1.5075 \\
                     &           & 21 &-& 5 & $7s6d^2$ & $^4$F$_{5/2}$ & 3.9777 \\
                     &           & 21 &-& 9 & $7s6d^2$ & $^4$P$_{1/2}$ & 1.0853 \\
                     &           & 21 &-& 11 & $7s6d^2$ & $^4$D$_{5/2}$ & 0.1932 \\
                     &           & 21 &-& 13 & $7s6d^2$ & $^2$F$_{5/2}$ & 0.4257 \\
                     &           & 21 &-& 14 & $7s6d^2$ & $^2$D$_{3/2}$ & 0.0793 \\
                     &           & 21 &-& 20 & $7s6d^2$ & $^4$D$_{5/2}$ & 0.2433 \\
    $7s7p6d$ & $^4$F$^{\rm o}_{7/2}$ &22 &-& 2 & $7s^26d$ & $^2$D$_{5/2}$ & 1.1232\\
                     &           & 22 &-& 5 & $7s6d^2$ & $^4$F$_{5/2}$ & 1.4187 \\
                     &           & 22 &-& 6 & $7s6d^2$ & $^4$F$_{7/2}$ & 3.8308 \\
                     &           & 22 &-& 7 & $7s6d^2$ & $^4$F$_{9/2}$ & 0.5212 \\
                     &           & 22 &-& 11 & $7s6d^2$ & $^4$D$_{5/2}$ & 0.4629 \\
                     &           & 22 &-& 13 & $7s6d^2$ & $^2$F$_{5/2}$ & 0.0015 \\
                     &           & 22 &-& 16 & $7s6d^2$ & $^4$G$_{7/2}$ & 0.0177 \\
                     &           & 22 &-& 19 & $7s6d^2$ & $^2$G$_{9/2}$ & 0.1330 \\
                     &           & 22 &-& 20 & $7s6d^2$ & $^4$D$_{5/2}$ & 0.2292 \\

    $7s7p6d$ & $^2$D$^{\rm o}_{5/2}$ &23 &-& 1 & $7s^26d$ & $^2$D$_{3/2}$ & 1.2505\\
                     &           & 23 &-& 2 & $7s^26d$ & $^2$D$_{5/2}$ & 2.4129 \\
                     &           & 23 &-& 4 & $7s6d^2$ & $^4$F$_{3/2}$ & 0.4658 \\
                     &           & 23 &-& 5 & $7s6d^2$ & $^4$F$_{5/2}$ & 1.7723 \\
                     &           & 23 &-& 6 & $7s6d^2$ & $^4$F$_{7/2}$ & 1.7920 \\
                     &           & 23 &-& 10 & $7s6d^2$ & $^4$P$_{3/2}$ & 1.6800 \\
                     &           & 23 &-& 11 & $7s6d^2$ & $^4$D$_{5/2}$ & 2.072 \\
                     &           & 23 &-& 13 & $7s6d^2$ & $^2$F$_{5/2}$ & 3.051 \\
                     &           & 23 &-& 14 & $7s6d^2$ & $^2$D$_{3/2}$ & 0.5276 \\
                     &           & 22 &-& 16 & $7s6d^2$ & $^4$G$_{7/2}$ & 1.0293 \\
                     &           & 23 &-& 20 & $7s6d^2$ & $^4$D$_{5/2}$ & 0.2203 \\
    
    $7s7p6d$ & $^4$D$^{\rm o}_{5/2}$ &25 &-& 1 & $7s^26d$ & $^2$D$_{3/2}$ & 0.4511 \\
                     &           & 25 &-& 2 & $7s^26d$ & $^2$D$_{5/2}$ & 0.8195 \\
                     &           & 25 &-& 4 & $7s6d^2$ & $^4$F$_{3/2}$ & 0.0587 \\
                     &           & 25 &-& 5 & $7s6d^2$ & $^4$F$_{5/2}$ & 0.6827 \\
                     &           & 25 &-& 6 & $7s6d^2$ & $^4$F$_{7/2}$ & 4.8072 \\
                     &           & 25 &-& 10 & $7s6d^2$ & $^4$P$_{3/2}$ & 0.7703 \\
                     &           & 25 &-& 11 & $7s6d^2$ & $^4$D$_{5/2}$ & 0.9500 \\
                     &           & 25 &-& 13 & $7s6d^2$ & $^2$F$_{5/2}$ & 0.1039 \\
                     &           & 25 &-& 14 & $7s6d^2$ & $^2$D$_{3/2}$ & 0.6897 \\
                     &           & 25 &-& 16 & $7s6d^2$ & $^4$G$_{7/2}$ & 0.1281 \\
                     &           & 25 &-& 20 & $7s6d^2$ & $^4$D$_{5/2}$ & 0.2901 \\
                     &           & 25 &-& 24 & $7s6d^2$ & $^4$G$_{7/2}$ & 0.2807 \\
\end{tabular}
\end{ruledtabular}
\end{table}

\begin{table}[b]
\caption{\label{tab:lifetimes}
Calculated lifetimes of the nine low-lying odd states of Ac.
Where available, experimental values of the transition frequency were used.
The assigned errors include the uncertainty in the calculated frequencies and $E1$ amplitudes.}
\begin{ruledtabular}
\begin{tabular}{rlcrrr}
\multicolumn{1}{c}{$N$}&
\multicolumn{2}{c}{State}&
\multicolumn{2}{c}{Energy (cm$^{-1}$)}&
\multicolumn{1}{c}{$\tau$ (ns)}\\  
\multicolumn{3}{c}{}&
\multicolumn{1}{c}{Exp.~\cite{NIST}}&
\multicolumn{1}{c}{Calc.}&
\multicolumn{1}{c}{}\\
\hline
3 & $7s^{2}7p$ & $^2$P$^{\rm o}_{1/2}$ &  & 7565 & 733(70)\\
8 & $7s^{2}7p$ & $^2$P$^{\rm o}_{3/2}$ &  & 12345 & 238(20) \\
12 & $7s7p6d$ & $^4$F$^{\rm o}_{3/2}$ & 13713 & 13958 & 317(30)\\
15 & $7s7p6d$ & $^4$F$^{\rm o}_{5/2}$ & 14941 & 15141 &  196(20)\\ 
18 & $7s7p6d$ & $^4$D$^{\rm o}_{1/2}$ & 17200 & 17049 &  139(14)\\
21 & $7s7p6d$ & $^4$D$^{\rm o}_{3/2}$ & 17736 & 17612 &  142(14)\\
22 & $7s7p6d$ & $^4$F$^{\rm o}_{7/2}$ & 17684 & 17715 &  385(40)\\
23 & $7s7p6d$ & $^4$D$^{\rm o}_{5/2}$ & 17951 & 18108 &  80(8)\\
25 & $7s7p6d$ & $^4$D$^{\rm o}_{5/2}$ &  & 18747 &   157(16)\\
\end{tabular}
\end{ruledtabular}
\end{table}

For the even state
$7s6d^{2}$~$^4$F$_{ 3/2}$
($N=4$ in Table~\ref{tab:Ac-energy}),
we calculate the lifetime to be
$4\E{-3}\,{\rm s}$.
We note, however, that the smallness of the energy interval between this state and the lower 
$7s^{2}7p$~$^2$P$^o_{ 1/2}$
state ($\omega\approx1000\,{\rm cm}^{-1}$) leads to instability in the calculation for the frequency. 
Therefore, without an experimental determination of the frequency, this should be considered an order-of-magnitude estimate.
The even states
$^2$D$_{ 5/2}$,
$^4$F$_{ 5/2}$,
$^4$F$_{ 7/2}$,
$^4$F$_{ 9/2}$, and
$^2$G$_{ 9/2}$, enumerated in Table~\ref{tab:Ac-energy} as 2, 5, 6, 7, and 19, respectively, are expected to be metastable.


\section{Conclusion}

We have performed accurate calculations of the energy levels, $g$-factors, transition matrix elements, and lifetimes for several of the low-lying states of neutral actinium using an all-order method based on the combination of the configuration interaction technique with the singles-doubles--coupled-cluster method.
Our calculations indicate good agreement with experiment for known levels, and we have provided predictions for many previously unidentified levels.
The calculations will help shed light on some potential misidentified levels in the literature, and will aid in the identification of new levels as experimental work continues.

\acknowledgments

This work was supported in part by the Australian Research Council.
BMR gratefully acknowledges financial support from Labex FIRST-TF.
The authors are grateful to Dmitry Budker and Marianna Safronova for pointing out
inconsistencies in tables of a previous version of this manuscript.
We express our gratitude to Dmitry Budker, Simon Rochester and Sebastian Raeder for discussions that stimulated this research.


\clearpage


\bibliography{Actinium} 

\end{document}